\documentclass{article}
\usepackage{amsmath}
\usepackage{amscd}
\usepackage{amsthm}
\usepackage{amssymb} 
\usepackage{latexsym}
\usepackage{euscript}
\usepackage{epsfig} 
\usepackage{graphics}
\usepackage{array}
\usepackage{enumerate}

\theoremstyle{definition}

\theoremstyle{remark}

\newcommand{\bel}[1]{\begin{equation}\label{#1}}

\newcommand{\be}{\begin{equation}}

\newcommand{\ba}{\begin{eqnarray}}
\newcommand{\ea}{\end{eqnarray}}
\newcommand{\rf}[1]{(\ref{#1})}
\newcommand{\bi}{\bibitem}
\newcommand{\qe}{\end{equation}}

\title{Randomness, chaos, and structure}
\author{Fatihcan Atay, Sarika Jalan, J\"urgen Jost\footnote{also: Santa Fe
Institute, USA;
jost@mis.mpg.de}\\
Max Planck
Institute for Mathematics in the Sciences, Leipzig, Germany}

\begin{document}
\maketitle

\begin{abstract}
We show how a simple scheme of symbolic dynamics distinguishes a chaotic 
from a random time series and how it can be used to detect structural 
relationships in coupled dynamics. This is relevant for the question at 
which scale in complex dynamics regularities and patterns emerge.
\end{abstract}

\section{Symbolic dynamics distinguishing chaotic from random dynamics}
\subsection{Random sequences and dynamical iterates} According to many 
popular accounts, chaotic dynamics seem to blur the distinction between 
determinism and randomness. While following a fixed rule, it is 
characteristic of chaotic dynamics that in the longer term no prediction 
of the iterates of given initial values is possible, and it therefore 
seems that sequences of points generated by chaotic dynamics are 
difficult, if not impossible, to distinguish from random sequences. Of 
course, this is not so, and one may exploit regularities in the 
relationships between subsequent points in the sequence to extract useful 
information about the underlying dynamics. By now, very sophisticated 
methods have been successfully developed, and we refer to \cite{KS} for a 
good account of the state of the art, describing both the older linear and 
the more recent non-linear tools, in particular phase space and other 
embedding methods, together with a rich spectrum of applications.\\ It is 
the purpose of the present article to analyze the relationship between 
randomness and chaos in an elementary manner using simple symbolic 
dynamics, and to utilize this to elucidate the formation of higher level 
structures through the coordination of lower level non-linear dynamics, as 
initiated in our earlier contribution \cite{AJ}.\\\\ The baseline 
situation is a sequence $x_n$, with $n \in {\mathbb N}$ as usual, of 
points randomly drawn from the unit interval $[0,1]$, independently of 
each other and all distributed according to the uniform density. The 
latter means that for each subinterval of $[0,1]$, the probability of 
finding $x_n$, for given $n$, in that interval is equal to its length.\\ 
We then consider the tent map
\ba
\label{1}
\nonumber
f&:&[0,1]\to [0,1]\\
f(x)&=&\begin{cases} 2x \ \ \ \ \ \text{ for } 0\le x \le 1/2 \\
 2-2x \text{ for } 1/2 \le x \le 1
\end{cases}
\ea
\begin{figure} 
\begin{center}
\includegraphics[scale=0.5]{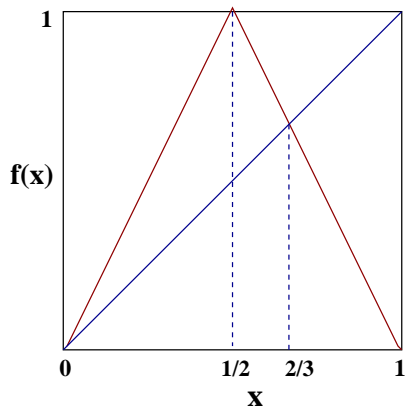}
\caption{{\small The graph of the tent map $f(x)$ has its peak at $x=1/2$ 
and intersects the diagonal at $x=2/3$.}}
\end{center}
\label{fig0}
\end{figure} 
as the basic example of a chaotic iteration
\bel{2}
x(n+1)=f(x(n)) \text{ for } n \in \mathbb{N}.
\qe
(In this paper, whenever we discuss a concrete map, $f$ will always be the 
tent map.)
Its stationary density $p$ on $[0,1]$ is the uniform density
\bel{3}
p(x)=1.
\qe
This means that if, for some generic initial value $x(0)$,\footnote{This 
qualification is needed because for all initial values of the form 
$x(0)=(1/2)^{-\nu}$ for some $\nu \in {\mathbb N}\cup \{0\}$, the 
iteration will end up in the fixed point 0. Those particular initial 
values, however, constitute a set of measure 0 in $[0,1]$ and can 
therefore be neglected for the purposes of our discussion.} we randomly 
choose $n \in {\mathbb N}$ and the corresponding point from the sequence 
$x(n)$, then again for each subinterval of $[0,1]$, the probability that 
the point lies in that subinterval is equal to its length.\footnote{It 
might seem that there is a profound difference between the ways the points 
are selected in the random and in the dynamical case. In the former one, 
we choose a random point for fixed ``time'' $n$, while in the latter one, 
we choose a point fixed by the dynamics at a random time $n$. The 
fundamental concept of ergodicity (which applies to our example), however, 
tells us that this leads to the same, that is temporal averaging is equal 
to spatial averaging.}

\subsection{Symbolic dynamics derived from time series}
\label{sd} 
We use the stationary density $p$ to construct derived symbolic dynamics 
according to the following rule, for some $a \in (0,1)$, 
\be \label{4} 
s(x)=\begin{cases}0 &\text{ if } 0 \le x \le a\\ 1 &\text{ if } a < x \le 
1. \end{cases} 
\qe 
So, from a sequence $x_n$ in $[0,1]$, one obtains a derived symbolic 
dynamics $s_n=s(x_n) \in \{0,1\}$. That sequence $x_n$ can now either be a 
random sequence chosen according to the density $p$, that is as above in 
our baseline situation, or a sequence $x(n)$ coming from our chaotic 
iteration \rf{2}.\\ 
The most natural choice for the partition point $a$ seems to be $1/2$. In 
that case, however, the symbolic dynamics does not distinguish between the 
random and the chaotic sequence. For our random sequence $x(n)$, when, 
say, $s_n=0$, then $0\le x(n) \le 1/2$, and each of the two subcases $0\le 
x \le 1/4$ and $1/4 <x \le 1/2$ occurs with probability $1/2$; in the 
first case, $s_{n+1}=0$, while in the second one, $s_{n+1}=1$, and so the 
two possible values for $s_{n+1}$ both occur with equal probability $1/2$. 
Since the same happens in case $s_n=1$, this is independent of the value 
of $s_n$, as for a random sequence. \\
The situation changes for other partition points $a$.\footnote{In 
\cite{Wa}, the difference between homogeneous partitions (based on the 
underlying Lebesgue measure) and generating partitions has been analyzed. 
In the present case, however, the partition point $a=1/2$ yields both a 
homogeneous and a generating partition.} The most significant and easy 
case is $a=2/3$ as the graph of the tent map intersects the diagonal 
there, see Figure 1; so, we consider
\be
\label{4a}
s(x)=\begin{cases}0 &\text{ if } 0 \le x \le 2/3\\
1 &\text{ if } 2/3 < x \le 1.
\end{cases}
\qe
For the random sequence, the values $s_n=0$ and $s_n=1$ occur 
independently with probabilities $2/3$ and $1/3$. For the chaotic 
sequence, when $s_n=1$, that is, $2/3 < x(n) \le 1$, the rule \rf{1} for 
the tent map yields $0\le x(n+1)< 2/3$, that is $s_{n+1}=0$. Thus, the 
successor of state one 1 is always state 0 for the chaotic map; no 
transition from 1 to 1 is possible. When we have the state $s_n=0$, both 
transitions are equally likely: when $0\le x(n) \le 1/3$, we have $0\le 
x(n+1) \le 2/3$, that is, $s_{n+1}=0$, while for $1/3 <x \le 2/3$, we get 
$s_{n+1}=1$. Thus, the state transition probabilities satisfy
\bel{5}
p(0|0)=p(1|0)=1/2,\ p(0|1)=1,\ p(1|1)=0
\qe
for the symbolic dynamics derived from the chaotic one while for the 
random one the probabilities $p(0)=2/3$ and $p(1)=1/3$ are independent of 
the previous state, that is,
\bel{5a}
p(0|0)=p(0|1)=2/3,\ p(1|0)=p(1|1)=1/3.
\qe

More generally, for a partition point $a$, the state probabilities are 
$p(0)=a, p(1)=1-a$. For the random dynamics with a uniform density, we 
have the transition probabilities
\begin{equation}
p(0|0) = p(0|1) = a, \text{ and } p(1|0)=p(1|1)=(1-a)
\label{rand-prob}
\end{equation}
For the tent map, we have
\begin{eqnarray}
p(0|0) = p(1|0) = \frac{1}{2}, \, \, p(1|1) = \frac{2 - 3a}{2(1-a)}, \, \, 
P(0|1)=\frac{a}{2(1-a)}   \, \, \text{ for } a < 2/3 \nonumber \\
p(0|0)=\frac{2 a -1}{a}, \, \, p(1|0) = \frac{1-a}{a}, \, \, p(1|1) = 0, \, \,
p(0|1)=1 \, \, \, \text{ for } a \ge 2/3.
\label{tent-prob}
\end{eqnarray}

\subsection{Symbolic dynamics from ordering relations between subsequent 
points}
The basic idea here has been first introduced by Bandt and Pompe 
\cite{BP}\footnote{even though they used it for a somewhat different 
purpose, as a method for approximating the entropy of a time series, 
instead of for a distinction between random and chaotic sequence as we 
shall do here} and is readily described by taking two points $x^1, x^2 \in 
[0,1]$ and the symbolic rule
\be
\label{6}
\nonumber
s(x^1,x^2)=\begin{cases}0& \text{ if } x^1 \le x^2\\
1& \text{ if } x^2 < x^1.
\end{cases}
\qe
We apply this to our random sequence, that is, at each step, we take 
$x^1=x_n, x^2=x_{n+1}$. Thus, we draw the points $x^1, x^2$ randomly and 
independently. The state probabilities are again $p(0)=p(1)=1/2$, but the 
transition probabilities become different:
\ba
\label{7}
\nonumber
p(0|0)=p(1|1)&=&1-\frac{1}{\sqrt{2}}\\
p(1|0)=p(0|1)&=&\frac{1}{\sqrt{2}}
\ea
because when $x_1$ is random, the average value of those $x_2$ with 
$x_1<x_2$ is $\frac{1}{\sqrt{2}}$. In fact, the symbolic dynamics is not 
Markovian since, for example $p(0|00)<p(0|01)$. More generally, the more 
0s have already occurred in sequel, the less likely it gets to observe 
another 0 as the next state. Thus, state probabilities depend on the 
entire past of the sequence. In particular, the symbolic sequence derived 
from our random sequence is not random itself.\\ The situation becomes 
simpler when we derive the symbolic dynamics from our chaotic map, 
$x^1=x(n), x^2=x(n+1)$. Here, the probabilities $p(0)$ and $p(1)$ are 
again equal, and the transition probabilities are given by \rf{5}, because 
$x\le f(x)$ precisely if $0\le x \le 2/3$. The process now is Markovian. 
For example, $p(0|00)=p(00|0)=p(000)=p(100)=p(00|1)=p(0|10)$:\footnote{We 
write the symbolic sequences here from left to right, that is, 10 means 
that we first see the symbol 1 and then the symbol 0; this explains the 
reversals of symbol order in these equations because conditioning is 
written from right to left.} three consecutive points $x(n), x(n+1), 
x(n+2)$ are between 0 and 2/3 precisely when $0\le x(n) \le 1/6$ while the 
symbolic sequence 100 occurs when $5/6 \le x(n) \le 1$. \\ The pattern 
becomes even more obvious when we consider three consecutive points 
$x^1,x^2,x^3$ and the symbol dynamics defined by
\ba
\label{8}
\nonumber
s(x^1,x^2,x^3)&=& 1\ \  \text{ if } x^1 < x^2 < x^3\\
\nonumber
s(x^1,x^2,x^3)&=& 2\ \  \text{ if } x^1 < x^2 > x^3\\
\nonumber
s(x^1,x^2,x^3)&=& 3\ \  \text{ if } x^1 > x^2 > x^3\\
s(x^1,x^2,x^3)&=& 4\ \  \text{ if } x^1 > x^2 < x^3.
\ea
(For simplicity, we neglect all cases of equality from now on because
those occur with probability 0.)\\
Regardless of how the points $x^1,x^2, x^3$ are drawn, the only possible 
transitions are 11, 12, 23, 24, 33, 34, 41, 42. When the points are 
randomly drawn, all of them occur. The transition probabilities are 
different, however: for example $p(1|1)<p(2|1)$. As before, the process is 
not Markovian in that case: for example $p(1|11)<p(1|14)$.\\ When the 
points are obtained from the chaotic iteration, $x^1=x(n), x^2=x(n+1), 
x^3=x(n+2)$, state 3 can no longer occur because we have already seen 
above that when $x(n+1)<x(n)$ we have $2/3 < x(n) <1$ and $0 < x(n+1)<1/3$ 
and therefore $x(n+2)>x(n+1)$. (In fact, the states 1 and 2 in the present 
dynamics correspond to the state 0 for the above symbolic dynamics 
obtained from the ordering between two consecutive points from a chaotic 
dynamics, while state 4 corresponds to state 1 in that latter dynamics.) 
Thus, this derived symbolic dynamics leads to an easy distinction between 
the random and the chaotic ones.

\subsection{Generalities}
The preceding makes possible a distinction between a particular chaotic 
iteration, the tent map, and a random iteration with the same underlying 
probability density. The question arises whether this symbolic method can 
also distinguish a more general class of chaotic iterations from a random, 
that is, to what extent this is useful for distinguishing chaos from 
randomness. One generalization is clear: the symbolic dynamics derived 
from ordering relations between consecutive points applies to any chaotic 
map that is conjugate to the tent map, like the logistic map. Of course, 
the stationary density will no longer be uniform in general, but it can 
readily be estimated from the time series produced by the dynamics, and 
one can take the random iteration based on that probability density for 
comparison. \\ 
For the symbolic dynamics derived from the partition, one should know the 
optimal partition point $a$; of course, when $a$ is unknown, one can try 
different ones so as to minimize the entropy of the resulting symbolic 
transition dynamics. For the random dynamics with a uniform density and 
partition point $a$, we had the state probabilities $p(0)=2/3, p(1)=1/3$ 
and independent transitions $p(i|j)$; this gives the entropy
\bel{9}
H=-\sum_{i=0,1}p(i)\sum_{j=0,1} p(j|i) \log p(j|i)=\log 3 -2/3\sim
.918.
\qe
For the tent iteration, we had $p(0|1)=1, p(1|1)=0,
p(1|0)=p(0|0)=1/2$, leading to the entropy
\bel{15}
H=2/3
\qe
which is significantly smaller. In general, we should expect that the 
symbolic dynamics for a chaotic map leads to a smaller entropy than for a 
random one (based on the same probability density). In fact, for the 
present example, the entropy difference between the random and the chaotic 
sequence is largest for $a=2/3$, see Figure 2. In contrast, for $a=1/2$ 
(which corresponds to the generating partition for the tent dynamics), the 
entropies are the same and the difference vanishes. When $a$ is close to 0 
or 1, the entropies of the random and of the chaotic sequence both become 
quite small, and therefore, the difference is likewise small. Our strategy 
5A is to choose such an $a$ that the difference is maximal so as to make 
the difference between random and chaotic dynamics most pronounced. \\
\begin{figure} \label{fig1} 
\begin{center}
\includegraphics[width=0.7\columnwidth]{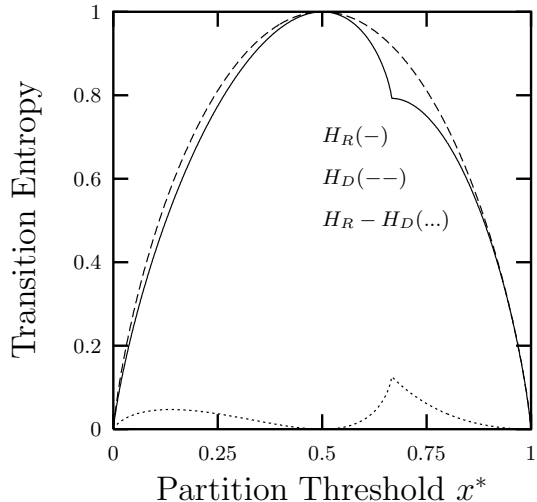}
\caption{{\small Transition entropies ($H_D$ and $H_R$) for the tent map 
and the corresponding random system, calculated from Eq.~(\ref{9}), by 
using (\ref{rand-prob}) and (\ref{tent-prob}) respectively.}}
\end{center}
\end{figure}

One can also view this in the following manner. When the baseline is some 
constant or similarly trivial dynamics, then the entropy difference 
between the chaotic tent map and the trivial map is largest for the value 
of $a$ that corresponds to the generating partition, that is, for $a=1/2$; 
in fact, one may consider this as the definition of the generating 
partition. This, however, then cannot distinguish between a deterministic 
chaotic iteration and a random sequence. When we want to find such a 
distinction, we should look rather for a partition where the entropy 
difference between those two sequences is maximized, and that lead us to 
the value $a=2/3$ in the present case. This is a very simple instance of 
the principle that interesting structure is neither trivial nor random.\\ 
While for more general chaotic dynamics, in general it is not easy, or 
perhaps even not possible, to find the optimal partition, still any 
partition that leads to an entropy difference between a chaotic iteration 
and a random sequence with the same underlying density yields detectable 
symbolic differences. Therefore, our method possesses some generality, and 
as an example, we have applied it to the H\'enon map in our recent work 
\cite{JJA}.

\section{Coupled dynamics}
In order to move beyond the simple comparison between a random and a 
chaotic sequence, we now consider coupled maps, as in \cite{AJ}. This 
means that we take some graph $\Gamma$, unweighted and undirected for 
simplicity, with $N$ vertices or nodes. Vertices $x,y$ of $\Gamma$ that 
are connected by an edge of $\Gamma$ are called neighbors, symbolically 
denoted by $x \sim y$. The number of neighbors of $x$ is denoted by $n_x$. 
For a parameter $\epsilon$, the coupling leads to the system
\bel{10}
x(n+1)= f(x(n))+ \frac{\epsilon}{n_x} \sum_{y \sim x} (f(y(n))-f(x(n))).
\qe
Thus, $x$ now adjusts its state not only the basis of its own present 
state, but also takes the state differences from its neighbors into 
account. The coefficients on the right hand side are chosen in such a 
manner that the total weight of all the contributions is 1, that is, the 
same as in \rf{2}. \\
Rewriting \rf{10} as
\bel{10a}
x(n+1)= (1-\epsilon)f(x(n))+ \frac{\epsilon}{n_x} \sum_{y \sim x} f(y(n))
\qe
leads to an alternative interpretation. Here, the node $x$ updates its 
state on the basis of a weighted average of a function of its own state 
and the corresponding values from its neighbors. In the special case where 
the graph $\Gamma$ is complete, that is, each of the $N$ vertices is 
connected with all $N-1$ other ones, when we then choose 
$\epsilon=\frac{N-1}{N}$, we obtain
\bel{10b}
x(n+1)=\frac{1}{N}\sum_{z \in \Gamma}f(z(n)).
\qe
Thus, in this particular case, the r.h.s. of the iteration dynamics 
equation is the same for all the vertices. Since then each of them updates 
its state not only by the same rule, but also with the same input, their 
states are all equal, that is $x(n+1)=y(n+1)$ for any two vertices $x,y$. 
Thus, the network is synchronized. It then turns out that synchronization 
also occurs for other values of the coupling strength $\epsilon$, see 
\cite{Kaneko84}, or for other graph topologies and is stable against 
perturbations, see e.g. \cite{JJ1}. \\ 
So, the conceptually simplest possibilities for the resulting network 
dynamics are:
\begin{enumerate}
\item The individual dynamics $x(n)$ are completely unrelated. This 
happens for $\epsilon=0$. In that case, each node behaves chaotically and 
is completely independent of the other ones.
\item The nodes synchronize, that is, $x(n)=y(n)$ for all nodes $x, y \in 
\Gamma$. In that case, the sum on the right hand side of \rf{10} becomes 
0, and consequently, each node behaves according to
\bel{11}
x(n+1)=f(x(n))
\qe 
which is the same as in the uncoupled case. 
\end{enumerate}
Thus, in both the uncoupled and the synchronized case, the individual 
dynamics are the chaotic ones given by \rf{2}. From looking at an 
individual node, we are not able to distinguish between the two scenarios. 
Both cases are extreme ones, and ultimately dynamically not very 
interesting, even though the synchronization of chaos after all is a 
surprising phenomenon. We therefore ask whether one can find and describe 
more interesting dynamics between those two extrema. At some level, such 
states should exhibit a behavior intermediate between the uncoupled and 
the fully synchronized dynamics. In \cite{AJ}, we described some emergent 
behavior on a longer time scale when transmission delays were introduced 
in \rf{10}. Here, we shall look for behavior that is intermediate 
regarding either the spatial coordination or the one of the state values 
$x(n)$. The paradigm for partial spatial coordination is the formation of 
dynamical clusters such that the nodes inside a cluster synchronize or 
otherwise coordinate their states, but that no such coordination occurs 
between clusters. An example of partial state value synchronization is the 
phase synchronization detected in \cite{JA} where the dynamical states 
$x(n)$ have their individual local temporal minima (or maxima) at the same 
times. \\ We shall now describe how those two types of dynamic behavior 
correspond to, and therefore can be detected by, certain types of derived 
symbolic dynamics according to our above scheme. We ask two questions:
\begin{enumerate}
\item In which settings or constellations do the symbolic dynamics derived 
from the state dynamics of some vertex exhibit regularities or 
characteristic features distinct from both the random \rf{5a} and the 
isolated chaotic one \rf{5}?
\item Under which circumstances, beyond the obvious one of 
synchronization, do the symbolic dynamics at different vertices show some 
correlations?
\end{enumerate}

\subsection{Local symbolic dynamics detecting collective properties of the 
dynamical system} We describe here three different types of relationships 
between local symbolic dynamics -- evaluated at a single node -- and 
collective properties of the dynamical system
\begin{enumerate}
\item {\bf Local symbolics and complexity of the collective dynamics:  
The largest Lyapunov exponent}\\
The Lyapunov exponents measure the rates of stretching or shrinking in a 
possibly high dimensional dynamical system. A positive Lyapunov exponent 
indicates an expanding, a negative one a contracting direction. A positive 
Lyapunov exponent is considered as an indication of chaos, and when there 
is more than one positive Lyapunov exponent, one speaks of hyperchaos. 
Lyapunov exponents are often difficult to compute in practice.\\ 
We have found that the transition probability for the symbolic rule 
\rf{4a} at any node of the dynamical network qualitatively matches the 
behavior of the largest Lyapunov exponent as becomes evident in Figure 
3.\footnote{In \cite{BP}, a qualitative similarity between the permutation 
entropy obtained from symbolic rules of the type \rf{6}, \rf{8} and the 
Lyapunov exponent of a time series derived from a single chaotic 
oscillator had been observed.}

\begin{figure}[h]
\begin{center}
\includegraphics[bb=56 350 504 781,width=10cm]{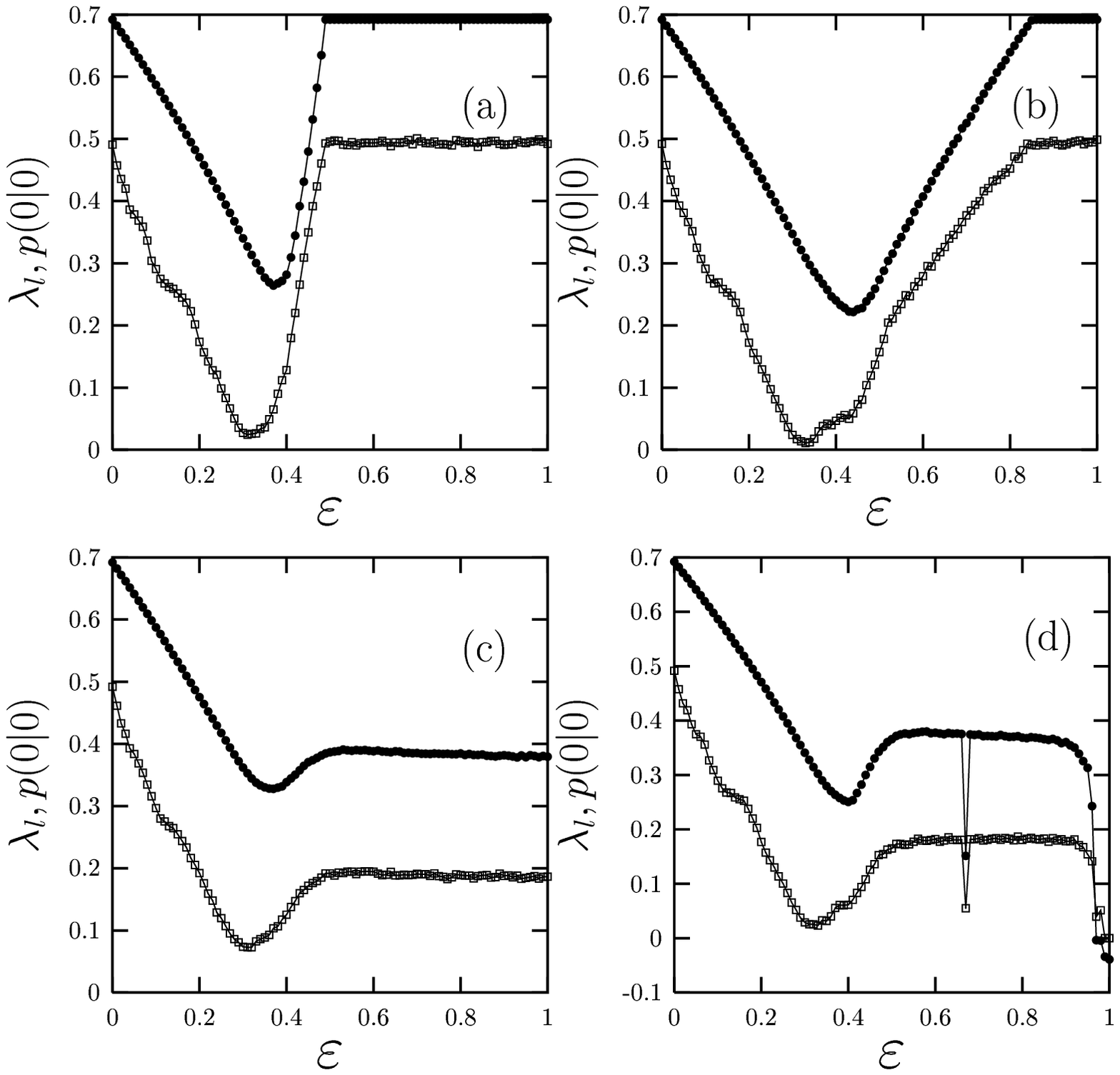}
\caption{{\small The largest Lyapunov exponent ($\bullet$) as a global 
measure of coupled dynamics and the transition probabilities $p(0|0) 
(\circ)$ of local symbolic dynamics, for various networks. The horizontal 
line represents coupling strengths. Figures are plotted for (a) a globally 
coupled network with $N=50$, (b) a scalefree network with $N=200$ and 
average degree 20, (c) and (d) for small world networks with $N=200$ and 
average degree 10 and 40 respectively.} } 
\end{center}
\label{fig2}
\end{figure} 
For the uncoupled 
tent map, we had $p(0|0)=1/2$, see \rf{5}, and it is remarkable that when 
the largest Lyapunov exponent decreases, this transition probability can 
even become 0, that is, two successive 0s no longer occur. The important 
point here is that some very easy measurement at one single node yields 
qualitative information about a global characteristic of the network that 
itself is difficult to compute. 
\item {\bf Local symbolics and coordination of individual dynamics in the 
network: Phase synchronization}\\ 
We say that two nodes $i,j$ are phase synchronized when the temporal 
maxima of $x^i(n)$ and $x^j(n)$ occur for the same values of $n$, that is, 
simultaneously; and we may require the same for the minima.\footnote{There 
exist different notions of phase synchronization in the literature, 
appropriate under different circumstances, see e.g. \cite{AK}. For our 
purposes, the one adopted here is most useful.} This is most easily 
detected by the symbolic dynamics \rf{8} because phase synchronization 
means that the symbols 2 and 4 for the corresponding symbolics occur 
simultaneously, and therefore also the other symbols by the transition 
constraints for \rf{8}. Phase synchronization is weaker than full 
synchronization, and therefore can occur more easily, that is for a wider 
range of coupling strengths and networks. It is a property of the state 
dynamics at a coarse level that may not be evident when focusing of the 
precise values of the states, that is, at the fine scale. Thus, the 
important point here is that the symbolics easily reveal a qualitative 
property at some coarse scale of the state dynamics. 
\\

\item {\bf Local symbolics and regularities on larger temporal and spatial 
scales:}\\
As explained in \cite{AJ}, the coupling, possibly in conjunction with 
transmission delays, may produce regularities at a longer time scale than 
accessible to the uncoupled individual dynamics that by their chaotic 
nature blur all distinctions on longer temporal scales. This must 
translate into a longer memory span of the symbolics. Conversely, memory 
effects, that is, long time correlations in the symbolics indicate a 
relevant longer temporal scale for the coupled dynamics.\\
Concerning larger spatial scales, it should be worth investigating the 
symbolic dynamics in hierarchical structures as investigated e.g. in 
\cite{Zh}.
\end{enumerate}

\subsection{Homogeneities at the symbolic level} 
The issue of phase synchronization just described can also be considered 
in the light of the second question raised above, namely the one about 
correlations in the symbolics. Obviously, when the network dynamics is 
synchronized, then so are the symbolics. But even when we do not have full 
synchronization, we should expect that the coupling leads to some 
coordination between the dynamics of the various nodes, and that should be 
detectable by suitable correlation measures. It is then natural to look at 
correlations between the symbolics, as asked above. Phase synchronization 
means that the symbolics of the different nodes become identical, but also 
the existence of dynamical regimes with weaker correlations between the 
symbolics is conceivable. It turns out that, remarkably, the transition 
probabilities for the symbolics at a single node can again give some 
indication of the degree of homogeneity of the symbolics across the 
network. That match, however, is not perfect; it works only for a certain 
range of values for the coupling strength $\epsilon$ for a given network.

\subsection{Symbolic dynamics as derived dynamics at a higher level of 
abstraction} 
Let us contemplate the general problem: The symbolic dynamics is derived 
from a lower level state dynamics and thus not autonomous. For the issue 
of emergence, it would be desirable that this dynamics at a higher level 
of abstraction develops at least some degree of autonomy, that is, that 
subsequent symbol values, or at least their probabilities, can be 
predicted from the values at previous times. For the probabilities, this 
is possible in the isolated case, see \rf{5}. The question remains whether 
this also emerges at the collective level.

\end{document}